\documentclass[10pt,conference]{IEEEtran} \newcommand{\figsza}{0.99} \newcommand{\figszb}{0.8}

\usepackage[cmex10]{amsmath}
\usepackage{amsfonts}
\usepackage{amssymb}
\usepackage{amsopn}
\usepackage{graphicx}
\usepackage{subfigure}
\usepackage{epsfig}
\usepackage{color}
\usepackage{setspace}
\usepackage{bbm}
\usepackage{algorithm}
\usepackage{algorithmic}

\newtheorem{result:opt}{Theorem}
\newtheorem{result:J}[result:opt]{Theorem}
\newtheorem{result:var:theta}[result:opt]{Theorem}
\newtheorem{result:var0}[result:opt]{Corollary}

\graphicspath{{fig/}}

\newcommand{\ud}{\,\mathrm{d}}

\newcommand{\tr}{\text{Tr }}
\newcommand{\bo}[1]{\boldsymbol{#1} } 
\newcommand{\tbo}[1]{\widetilde{\boldsymbol{#1}}}

\newcommand{\bs}{\begin{smallmatrix}}
\newcommand{\es}{\end{smallmatrix}} 

\makeatletter
\def\blfootnote{\xdef\@thefnmark{}\@footnotetext}
\makeatother

\begin{document}


\title{Controlled Collaboration for Linear Coherent Estimation in Wireless Sensor Networks}

\author{\IEEEauthorblockN{Swarnendu Kar and Pramod K. Varshney}
\IEEEauthorblockA{Dept. of Electrical Engineering and Computer Science, Syracuse University, Syracuse, NY,  13244, USA\\
Email: \{swkar,varshney\}@syr.edu}
}

\maketitle

\begin{abstract}
We consider a wireless sensor network consisting of multiple nodes that are coordinated by a fusion center (FC) in order to estimate a common signal of interest. In addition to being coordinated, the sensors are also able to collaborate, i.e., share observations with other neighboring nodes, prior to transmission. In an earlier work, we derived the energy-optimal collaboration strategy for the single-snapshot framework, where the inference has to be made based on observations collected at one particular instant. In this paper, we make two important contributions. Firstly, for the single-snapshot framework, we gain further insights into partially connected collaboration networks (nearest-neighbor and random geometric graphs for example) through the analysis of a family of topologies with regular structure. Secondly, we explore the estimation problem by adding the dimension of time, where the goal is to estimate a time-varying signal in a power-constrained network. To model the time dynamics, we consider the stationary Gaussian process with exponential covariance (sometimes referred to as Ornstein-Uhlenbeck process) as our representative signal. For such a signal, we show that it is always beneficial to sample as frequently as possible, despite the fact that the samples get increasingly noisy due to the power-constrained nature of the problem. Simulation results are presented to corroborate our analytical results. 
\end{abstract}

\section{Introduction}
We consider a wireless sensor network deployed for the purpose of monitoring a common phenomenon. The sensors transmit their observations to a fusion center in a cooperative manner, so as to conserve the overall (energy/power) resources available in the network.\blfootnote{This research was partially supported by the National Science Foundation under Grant No. 0925854 and the Air Force Office of Scientific Research under Grant No. FA-9550-10-C-0179.} In the widely researched area of \emph{distributed estimation} \cite{Rib06},\cite{Cui07}, the sensors are coordinated to ensure that, without communicating with one-another, they collectively maximize the quality of inference at the FC. In the amplify-and-forward approach to distributed estimation, the sensors linearly scale (based on the energy allocated) their observations while communicating with the FC. Since no coding across time is required and no non-linear processing is required at the nodes, amplify-and-forward techniques are operationally simple and enjoy widespread usage in the literature. Early application of the amplify-and-forward technique in distributed estimation was explored in \cite{Cui07} and \cite{Xiao08}, where orthogonal and coherent  multiple access channels (MAC) were considered. In an orthogonal MAC setting, each sensor has its own channel for communication with the FC while in the coherent MAC scenario, all the sensors coherently form a beam into a common channel which is then received by the FC. Other approaches to distributed estimation consider rate constraint in transmission, where the sensor nodes are required to quantize their observations before transmission to the FC, examples include \cite{Rib06} and more recently \cite{KarTSP12}.  

Some recent studies \cite{Fang09},\cite{KarISIT12} have demonstrated significant improvement over the distributed framework by allowing the sensor nodes to share their observations with other neighboring nodes prior to transmission to the FC. This act of sharing observations is referred to as \emph{spatial collaboration}. In an orthogonal MAC setting with a fully connected network, it has been shown in \cite{Fang09} that it is optimal to compute the estimates in the network and use the best available channel to transmit the estimated parameter. In a recent work that considered a coherent MAC channel for communication with the FC \cite{KarISIT12}, we presented an extension of the amplify-and-forward framework that allowed spatial collaboration in a partially connected network topology. It was observed that even a sparsely connected network was able to realize a performance which was very close to that of a fully connected network. This is due to the fact that in an amplify-and-forward framework, the observation noise is also amplified along with the signal, thereby significantly increasing the energy required for transmission. Spatial collaboration, in effect, smooths out the observation noise, thereby improving the quality of the signal that is transmitted to the FC using the same energy resources.  

In this paper, we explore the potential of collaborative estimation further by making two significant contributions. First, though it was observed earlier that even a moderately connected network performs almost as well as a fully connected network, no analytical results were presented. In this paper, we extend our previous work by analyzing the estimation performance for partially connected collaboration networks. Though the analysis of arbitrary network topologies is a difficult problem, we derive the estimation performance for a family of structured network topologies, namely the $Q$-cliques. We demonstrate that the insights obtained from the structured topology apply approximately to two practical topologies, namely the nearest neighbor and random geometric graphs, of similar connectivity. Given a particular network topology, we investigate two different energy allocation schemes for data transmission. In addition to the optimal energy-allocation (EA) scheme as derived in \cite{KarISIT12}, we also consider the suboptimal but easy-to-implement equal energy-allocation scheme, where neighboring observations are simply averaged to mitigate the observation noise. For both of these schemes, we derive the performance in a closed form in the asymptotic domain where the number of nodes is large and the overall transmission capacity of the network is held constant. These results offer insights into the relationship between estimation performance and problem parameters like channel and observation gains, prior uncertainty and extent of spatial collaboration.

The collaborative estimation problem has so far been analyzed in the single-snapshot context, where energy-constrained spatial sampling is performed at one particular instant and the inference is made using those samples. In our second contribution in this paper, we extend the problem formulation to consider power-constrained inference of a random process, where the goal is to estimate the process for \emph{all} time instants. In contrast to the simple snapshot framework, this involves obtaining multiple samples in time and computing the filtered estimates for any desired time instants, including time instants where observation samples are not available. Since collection of each sample involves the expenditure of energy resources, the appropriate constraint in this situation is the energy spent per unit time (or power). A key parameter here is the sampling frequency, the choice of which affects the overall estimation performance. Note that a higher sampling frequency usually means that one can better capture the temporal variations. However, with a power constraint, less energy is available for the collection of each of those samples, which would result in more noisy samples. This trade-off is investigated in the context of a Gaussian random process with exponential covariance, where it turns out that a higher sampling frequency \emph{always} results in better estimates.

\begin{figure}[htb]
\centering
    \includegraphics[width=\figsza \columnwidth]{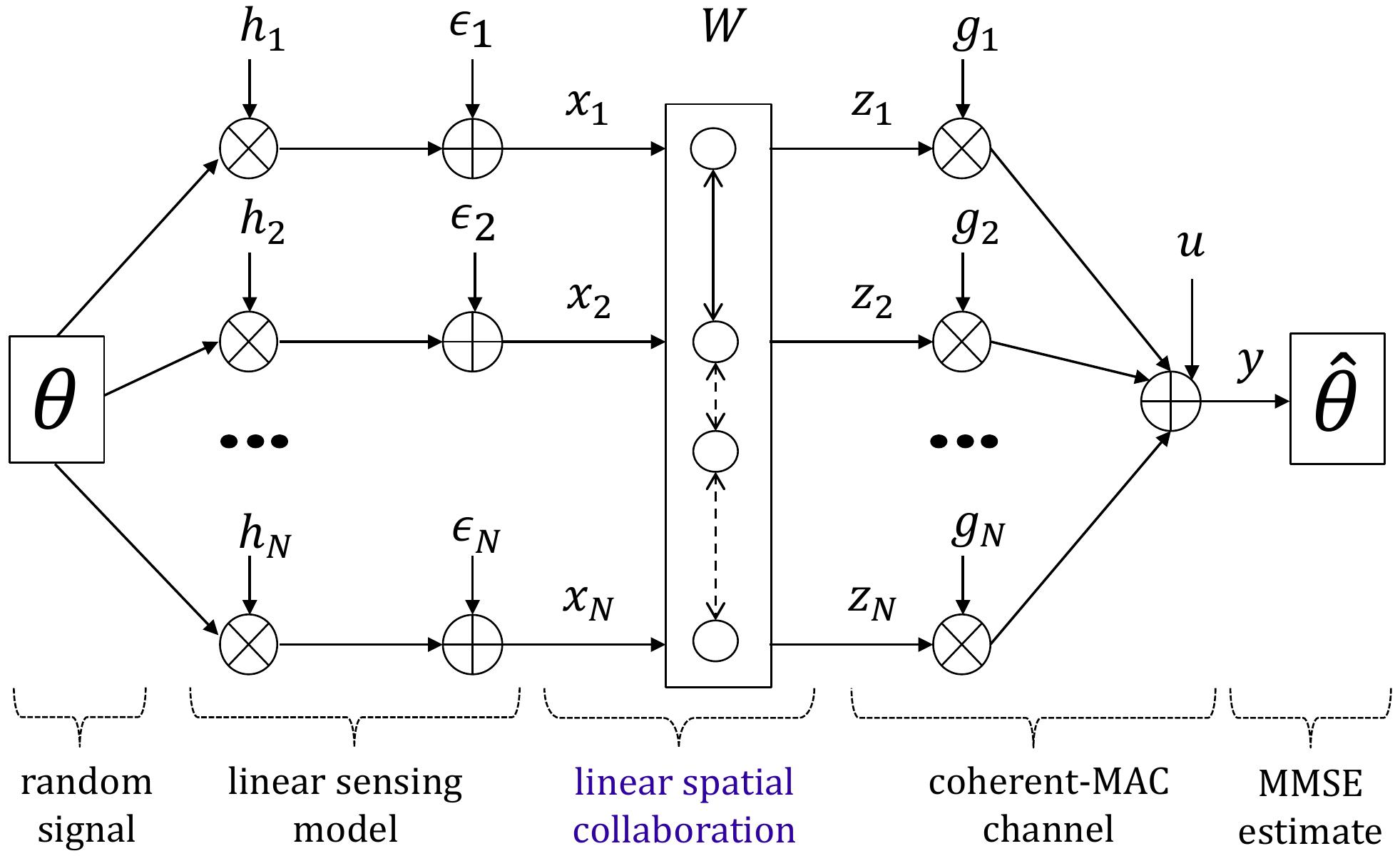}
  \caption{Wireless sensor network performing collaborative estimation.}
  \label{fig:schematic}
\end{figure}

\section{Problem Formulation}
We first consider the single snapshot estimation problem, where the dimension of time is ignored. We will extend the discussion to time varying Gaussian process later in Section \ref{sec:prob:form:OU}. The single snapshot framework is depicted in Figure \ref{fig:schematic}. The parameter to be estimated, $\theta$ (written without any time subscript to signify the single snapshot nature of the problem), is assumed to be a zero-mean Gaussian source with prior variance $\eta^2$. Different noisy versions of $\theta$ are observed by $N$ sensors. The observation vector is $\bo x=[x_1,\ldots,x_N]$ where $x_n=h_n \theta+\epsilon_n$, with $h_n$ and $\epsilon_n$ denoting the observation gain and measurement noise respectively. The measurement noise variables $\{\epsilon_n\}_{n=1}^N$ are assumed to be independent and identically distributed (iid) Gaussian random variables with zero mean and variance $\sigma^2$. 

Let the availability of collaboration links be represented by the adjacency matrix $\bo A$, where $A_{nm}=1$ (or $A_{nm}=0$) implies that node $n$ has (or does not have) access to the observation of node $m$. Define an $\bo A$-sparse matrix as one for which non-zero elements may appear only at locations $(n,m)$ for which $A_{nm}=1$. The set of all $\bo A$-sparse matrices is denoted by $\mathcal S_A$. Corresponding to an adjacency matrix $\bo A$ and an $\bo A$-sparse matrix $\bo W$, we define \emph{collaboration} in the network as individual nodes being able to linearly combine local observations from other collaborating nodes
\begin{align}
z_n=\sum_{m\in \mathcal A(n)} \bo W_{nm}x_m, \label{def:zn}
\end{align}
where $\mathcal A(n)\triangleq\left\{m:\; \bo A_{nm}=1 \right\}$, without any further loss of information. In effect, the network is able to compute a one-shot spatial transformation of the form $\bo z=\bo W \bo x$. 
In practice, this transformation is realizable when any two neighboring sensors are close enough to ensure reliable information exchange. Note that, when $\bo W$ is restricted to be diagonal (in other words, when $\bo A=\bo I$), the problem reduces to the amplify-and-forward framework for distributed estimation, which is widely used in the literature \cite{Cui07},\cite{Xiao08},\cite{Fang09} due to its simplicity in implementation and provably optimal information theoretic properties for simple networks \cite{Gastpar03}.

The transformed observations $\{z_n\}_{n=1}^N$ are transmitted to the FC through a coherent MAC channel, so that the received signal is $y=\bo g^T \bo z +u$, where $\bo g$ and $u$  describe the channel gains and the channel noise respectively. The channel noise $u$ is assumed to be Gaussian distributed with zero mean and variance $\xi^2$. The FC receives the noise-corrupted signal $y$ and computes an estimate of $\theta$. Since $y$ is a linear Gaussian random variable conditioned on $\theta$, 
\begin{align}
\begin{split}
\theta &\sim \mathcal N(0,\eta^2),\mbox{ and} \\
y|\theta &\sim \mathcal N \left(\underbrace{\bo g^T \bo W \bo h}_{\bs \triangleq \mu \; \text{(net gain)} \es} \theta,\, \; \underbrace{\bo g^T \bo W \bo \Sigma \bo W^T \bo g +\xi^2}_{\bs \triangleq \zeta^2 \; \text{(net noise variance)} \es}\right),
\end{split} \label{linear:model}
\end{align}
the minimum-mean-square-error (MMSE) estimator $\widehat \theta=\mathbb E\left[ \theta|y \right]$ is the optimal fusion rule. From estimation theory (for details the reader is referred to \cite{Kay93}), the MMSE estimator and resulting distortion $D_{\bo W}$ is given by
\begin{align} 
\widehat \theta &=\frac{1}{1+\frac{\zeta^2}{\eta^2 \mu^2}} \frac{y}{\mu},\mbox{ and } \frac{1}{D_{\bo W}}=\frac{1}{\eta^2}+J_{\bo W}, \; J_{\bo W}=\frac{\mu^2}{\zeta^2}, \label{JW}
\end{align}
where the quantity $J_{\bo W}$ is the Fisher Information and $\mu$ and $\zeta^2$ are the net gain and net noise variance as defined in Equation \eqref{linear:model}. The cumulative transmission energy required to transmit the transformed observations $\bo z$ is
\begin{equation} \begin{aligned}
\mathcal E_{\bo W}&=\mathbb E[\bo z^T \bo z]=\tr\left[\bo W \bo E_{\textsf x} \bo W^T\right], \mbox{ where} \\
\bo E_{\textsf x} &\triangleq \mathbb E[\bo x \bo x^T] = \eta^2 \bo h\bo h^T+\bo \Sigma. \label{def:PW}
\end{aligned} \end{equation}

\subsection{Collaboration strategies}
Note that the quantities $\mu$, $\zeta^2$ and, therefore, the distortion $D$ (equivalently $J$) and also the transmission energy $\mathcal E$ depend on the choice of the collaboration matrix $\bo W$. As indicated earlier, we explore two strategies to determine $\bo W$, namely 1) optimal and  2) equal energy-allocation (EA) strategies, that stem from two different engineering considerations. 

In the optimal EA strategy, we assume that the FC knows the channel and observation gains and also the collaboration topology precisely. In such a situation, the FC can compute the optimal collaboration matrix subject to a cumulative transmission energy constraint
\begin{align}
\mbox{(Optimal EA)}\quad \bo W_\textsf{opt}= \arg \min_{\bo W \in \mathcal S_A} D_{\bo W}, \; \mbox{ s.t. } \mathcal E_{\bo W} \le \mathcal E, \label{EA:opt}
\end{align}
and communicate the corresponding weights $\bo W_\textsf{opt}$ to the sensor nodes via a separate and reliable control channel. The exact form of $\bo W_\textsf{opt}$ and corresponding $J_\textsf{opt}$ were derived in \cite{KarISIT12} and are briefly described as follows.

\begin{result:opt}[Optimal single-snapshot estimation, \cite{KarISIT12}] \label{result:opt:lbl}
Let $L$ be the cardinality of $\bo A$, which is also the number of non-zero collaboration weights. In an equivalent representation, construct $\bo w \in \mathbb R^L$ by concatenating those elements of $\bo W$ that are allowed to be non-zero. Accordingly, define the $L\times L$ matrix $\bo \Omega$ and $L\times N$ matrix $\bo G$ such that the identities
\begin{align}
\tr\left[\bo W \bo E_{\textsf x} \bo W^T\right]=\bo w^T \bo \Omega \bo w, \mbox{  and  }\bo g^T \bo W =\bo w^T \bo G,
\end{align}
are satisfied. Then the optimal Fisher Information is,
\begin{align}
\begin{split}
&J_{\textsf{opt}}=\bo h^T\left(\bo \Sigma+\bo \Gamma/\mathcal E_\xi  \right)^{-1} \bo h, \mbox{ where} \\
&\; \mathcal E_\xi \triangleq \mathcal E/\xi^2, \mbox{ and } \bo \Gamma \triangleq \left(\bo G^T \bo \Omega^{-1}\bo G\right)^{-1},
\end{split} \label{J:opt:actual}
\end{align}
which is achieved when the collaboration weights are $\bo w_{\textsf{opt}}=\kappa \bo \Omega^{-1}\bo G \bo \Gamma\left(\bo \Sigma+ \bo \Gamma/\mathcal E_\xi  \right)^{-1} \bo h$, with the scalar $\kappa$ chosen to satisfy $\bo w_{\textsf{opt}}^T\bo \Omega \bo w_{\textsf{opt}}=\mathcal E$. $\bo W_\textsf{opt}$ is the matrix equivalent of $\bo w_\textsf{opt}$.
\end{result:opt}

When either the FC is computationally limited or reliable control channels are not available, the optimal EA strategy cannot be implemented. In these situations, one reasonable way of assigning transmission energy and collaboration weights at each node may be the equal EA strategy, where all the sensors are allocated equal transmission energy (namely $\frac{\mathcal E}{N}$). In addition, the $n$th sensor equally weighs all the observations from its neighbors where the weights (say $\{d_n\}_{n=1}^N$) are chosen to satisfy $\mathbb E\left[ z_n^2 \right]=\frac{\mathcal E}{N}$. Note from \eqref{def:zn} that $z_n= d_n\sum_{m\in \mathcal A(n)} (h_m\theta+ \epsilon_m)$. Consequently,
\begin{equation} \begin{aligned}
&\mbox{(Equal EA)} \quad \left[\bo W_{\textsf{eq}}\right]_{nm}=\left\{ \begin{array}{rl}
d_n,              & \mbox{if } m\in \mathcal A(n) \\
0,                  & \mbox{else}
\end{array} \right.,\\
&\quad d_n=\sqrt{\frac{\mathcal E/N}{ \left(\sum_{m\in \mathcal A(n)} h_m \right)^2\eta^2 +| \mathcal A(n) | \sigma^2}},
\end{aligned} \label{EA:eq} \end{equation}
where $| \mathcal A(n) |$ denotes the number of neighbors of $n$. The Fisher Information corresponding to the equal EA strategy is simply $J_\textsf{eq}\triangleq J_{\bo W_\textsf{eq}}$, which can be obtained by applying Equation \eqref{JW}.

Once the collaboration strategy (namely, either optimal or equal EA) is chosen and a cumulative operating energy $\mathcal E$ is specified, the resulting distortion performance (FI-s $J_\textsf{opt}$ or $J_\textsf{eq}$) depends on the following problem parameters, 1) signal prior, measurement noise and channel noise, which were assumed to be Gaussian distributed with zero mean and variances $\eta^2$, $\sigma^2 \bo I_N$ and $\xi^2$ respectively, 2) observation and channel gains, and 3) the collaboration topology. To obtain analytical expressions for FI-s, it is clear that we need to make further simplifying assumptions on the observation/channel gains (which will be discussed in Section \ref{sec:snapshot}) and also the topology for collaboration.

\begin{figure}[htb]
\centering
    \includegraphics[width=\figszb \columnwidth]{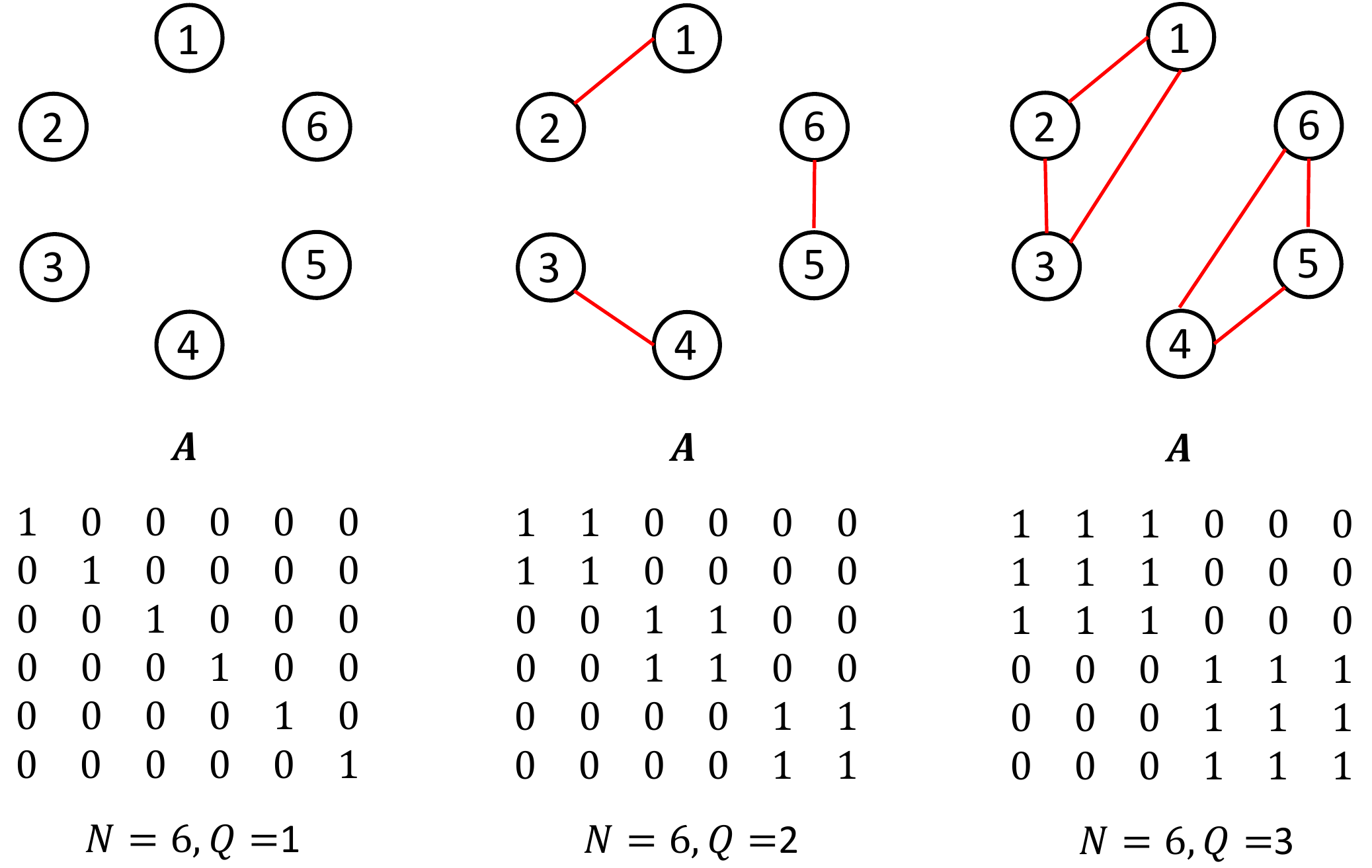}
  \caption{Example of $Q$-cliques.}
  \label{fig:Qconn}
\end{figure}

\subsection{Partially connected networks} \label{sec:partial}
\begin{figure}[htb]
\centering
\subfigure[Random geometric graph with radius $r=0.2$ (total 44 links)]{
\includegraphics[width=\figszb \columnwidth]{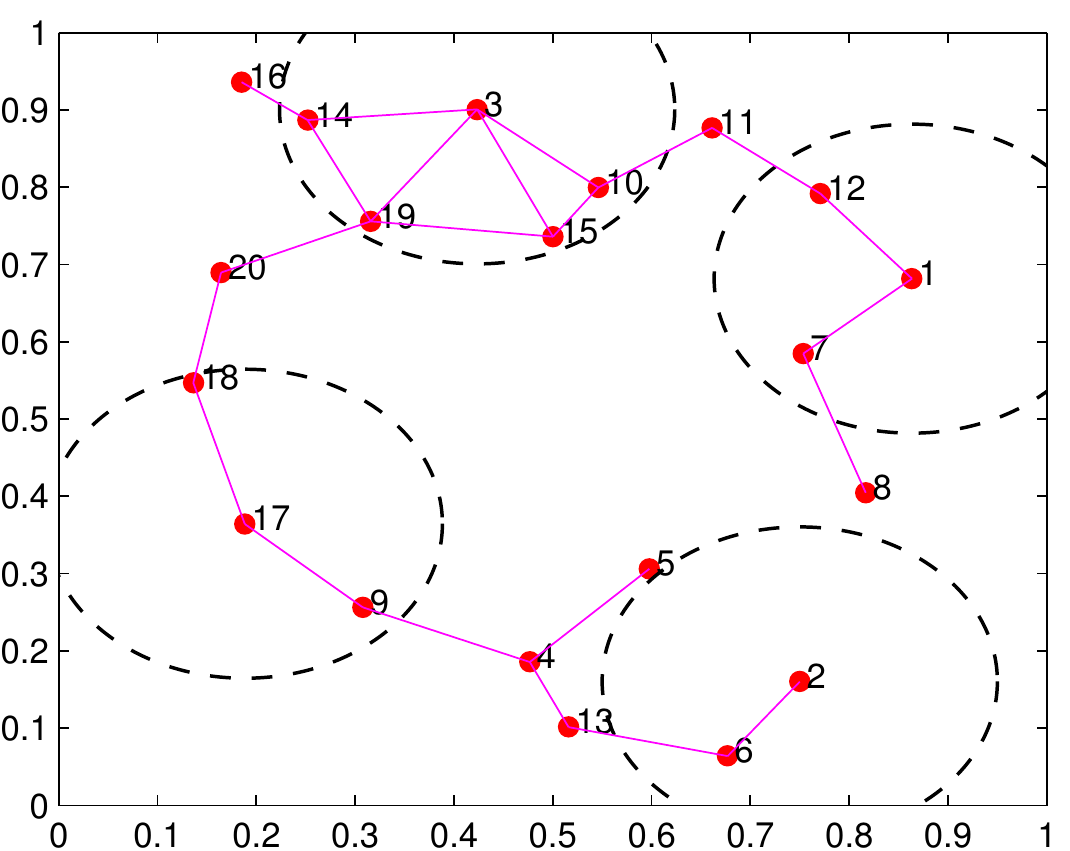}
\label{fig:topology:rgg}
}
\subfigure[Nearest neighbor topology with $Q=3$ (total 40 links) ]{
\includegraphics[width=\figszb \columnwidth]{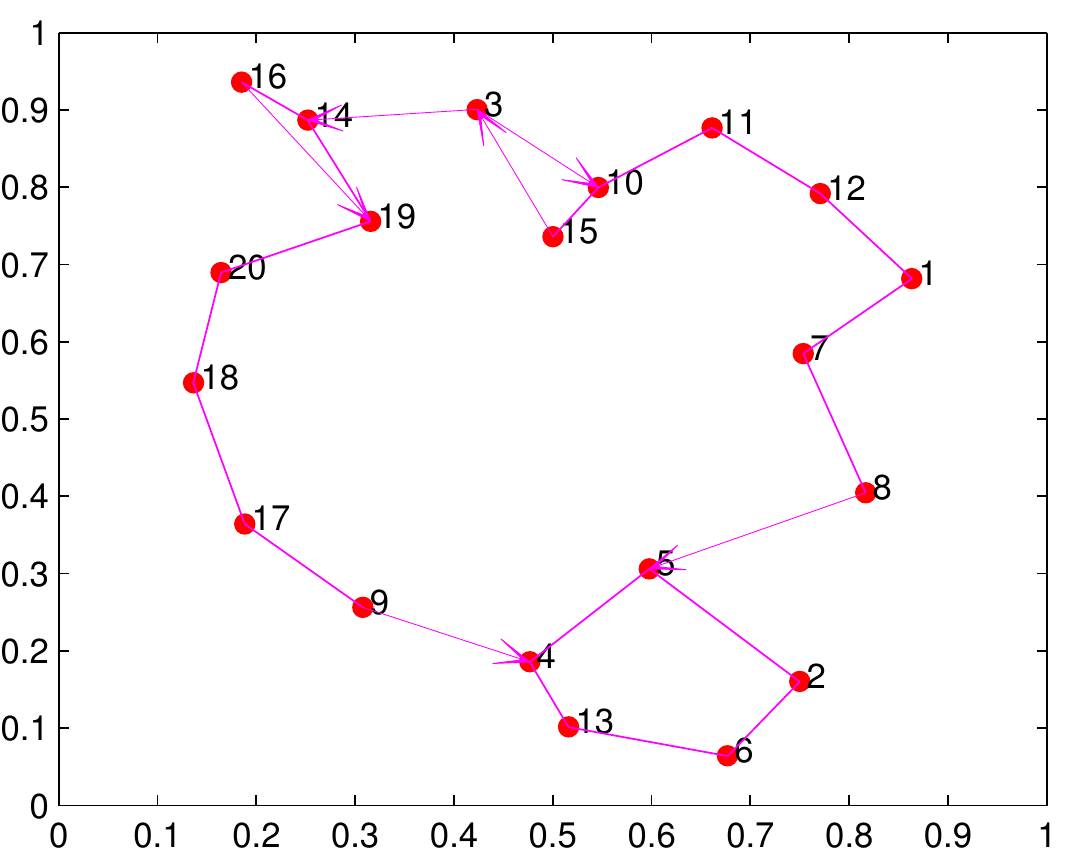}
\label{fig:topology:nn}
}
\caption{Example of two practical collaboration topologies for a $N=20$-node network. Bidirectional links are shown without arrows.}
\label{fig:topology}
\end{figure}

With the goal to investigate partially connected collaboration topologies, we adopt the following methodology. Intuitively, we expect the distortion to decrease as the network becomes more connected (since it adds more degrees of freedom) and it is our aim to obtain asymptotic limits that explicitly reflect the effect of \emph{connectedness}. Since the analysis of arbitrary topologies is difficult, we derive our analytical results for a structured collaboration topology that consists of several fully-connected clusters (or cliques) of finite size $Q$, as illustrated in Figure \ref{fig:Qconn}. To be precise, if $N=KQ$, we have $\bo A=\bo I_K \otimes \left( \bo 1_Q \bo 1_Q^T \right)$. Since all of the nodes in a $Q$-clique network are $(Q-1)$-connected, the performance of this special topology may serve as an approximation to other topologies where the average number of neighbors per node is $(Q-1)$. To demonstrate the efficacy of this approximation, we will compare the analytical results for $Q$-clique networks with numerical results for two practical collaboration topologies, namely 1) nearest-neighbor (NN) topology and 2) random geometric graphs (RGG) \cite{Freris10}. For the NN-topology, a sensor receives the observations from its nearest $Q-1$ neighbors. For the RGG topology, a sensor collaborates with all other sensors that are located within a circle of radius $r$ with the sensor at the center. The expected number of neighbors, which is a function of $r$, can be derived using geometric arguments, thereby enabling comparison with an equivalent $Q$-clique topology. Examples of these two topologies are illustrated in Figures \ref{fig:topology:rgg} and \ref{fig:topology:nn} for a network with $N=20$ nodes. It may be noted that unlike the NN topology, all collaboration links of an RGG topology are bidirectional by definition. 

\subsection{Ornstein-Uhlenbeck process } \label{sec:prob:form:OU}
We now bring in the dimension of time, and consider the problem of power-constrained estimation of time-varying signals. In order to model the temporal dynamics, we assume that the signal of interest is a stationary zero-mean Gaussian random process $\theta_t$ with exponential covariance function
\begin{align}
\mathbb E\left[ \theta_{t_1}, \theta_{t_2} \right]=\eta^2 e^{-\left( |t_1-t_2| \right)/\tau}, \label{cov:OU}
\end{align}
where $\eta^2$ and $\tau$ represent the magnitude and temporal variation of the parameter respectively. Note that $\tau\rightarrow 0$ implies that the signal changes very rapidly, while $\tau\rightarrow \infty$ means that the signal is constant over time. Such a process (with covariance parameterized by $\eta^2$ and $\tau$) is widely used in the literature \cite{Niu10} due to its ability to model a time varying Gaussian process while providing a relatively simple framework for analysis. This process is also sometimes known as the Ornstein-Uhlenbeck (OU) process. 

\begin{figure}[htb]
\centering
    \includegraphics[width=\figsza \columnwidth]{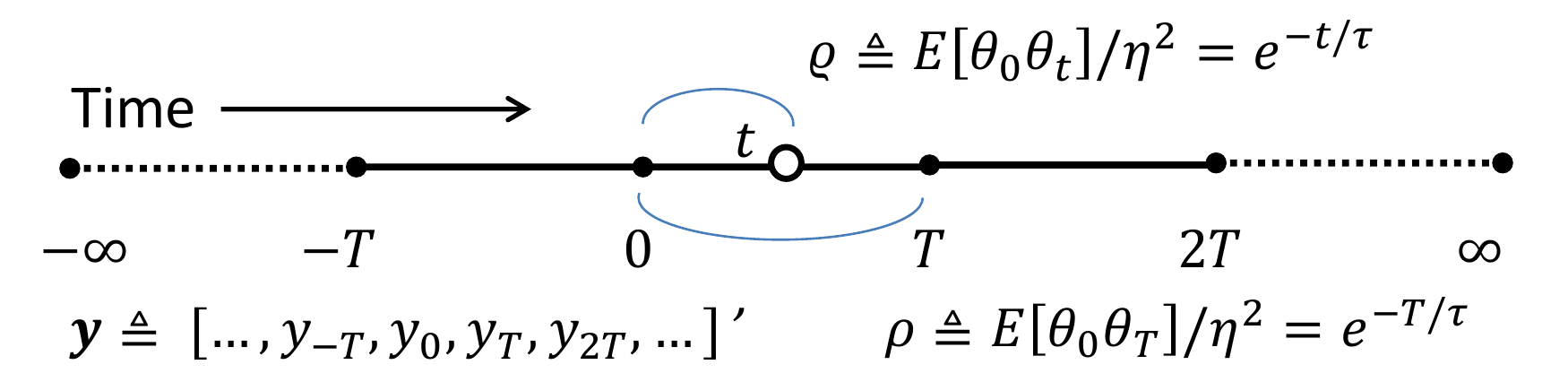}
  \caption{Periodically sampled Ornstein-Uhlenbeck process.}
  \label{fig:sampling}
\end{figure}

Let $P$ denote the power constraint in the network. We assume that the OU process is sampled periodically with the period $T$ (see Figure \ref{fig:sampling}), which implies that a total of $\mathcal E=PT$ energy units is available for each sampling instant. Let the spatial sampling at each instant be performed in a manner similar to the single-snapshot framework discussed earlier, namely through collaboration and coherent amplify-and-forward beamforming. Let the signal received by the FC at instant $t$ be denoted as $y_t$, so that the entire observed sequence can be represented by the following infinite-dimensional vector $\bo y\triangleq [\ldots,y_{-T}, y_0, y_T,y_{2T},\ldots ]'$. From the linear Gaussian model $y|\theta$ in \eqref{linear:model} and subsequent description of Fisher Information $J$ in \eqref{JW}, the spatial sampling process can be abstracted (via an appropriate scaling)  through the additive model $y_t=\theta_t+v_t$, where the \emph{aggregate} noise $v_t\sim \mathcal N\left(0,\frac{1}{J}\right)$ summarizes the uncertainty due to $\left\{ \epsilon_{n,t} \right\}_{n=1}^N$ (measurement noise at sensors) and $u_t$ (channel noise). We assume $\epsilon_{n,t}$ and $u_t$ to be temporally white, from which it follows that $v_t$ is temporally white as well. In vector notations,
\begin{align}
\bo y=\bo \theta +\bo v, \quad \bo v\sim \mathcal N \left(0,\frac{1}{J} \bo I\right), \label{y:theta:v}
\end{align} 
where $\bo \theta \triangleq [\ldots,\theta_{-T}, \theta_0, \theta_T,\theta_{2T},\ldots]'$ and $\bo v \triangleq [\ldots,v_{-T}, v_0, v_T, v_{2T},\ldots ]'$. Having observed $\bo y$, the MMSE estimator of $\theta_t$ (the value of OU process at any instant $t$) is given by the conditional expectation (refer to \cite{Kay93} for details)
\begin{align}
\widehat \theta_t=\mathbb E[\theta_t|\bo y] = \bo R_{\theta_t \bo y'} \bo R_{\bo y \bo y'}^{-1} \bo y, \label{E:theta:t}
\end{align}
where $\bo R_{\theta_t \bo y'}\triangleq \mathbb E\left[ \theta_t \bo y'\right]$ and $R_{\bo y \bo y'}\triangleq \mathbb E\left[ \bo y \bo y'\right]$. Moreover, the variance of $\widehat \theta_t$ is given by
\begin{align}
\text{Var}\left(\theta_t|\bo y\right) = \eta^2-\bo R_{\theta_t \bo y'} \bo R_{\bo y \bo y'}^{-1} \bo R_{\theta_t \bo y}. \label{var:theta:t}
\end{align}
Since $y_t$ is sampled periodically at instants $t=kT,\, k\in \mathbb Z$ and $\bo y$ contains infinite elements in both time directions, the conditional variance $\text{Var}\left(\theta_t|\bo y\right)$ is also expected to be periodic in time, i.e., $\text{Var}\left(\theta_t|\bo y\right)=\text{Var}\left(\theta_{t+kT}|\bo y\right),\; \forall t,k$. Hence any interval of length $T$, say $t\in[0,T]$ is sufficient for analyzing the conditional variance \eqref{var:theta:t}.  Since we are interested in estimating the OU process at \emph{all} time instants, the quality of inference has to be summarized by a metric that is independent of time $t$. We consider two such performance metrics, first of which is the average variance 
\begin{align}
\text{Avar}(T) \triangleq \frac{1}{T}\int_0^T \text{Var}\left(\theta_t|\bo y\right) \ud t. \label{def:avar}
\end{align}
Note that average variance depends on the sampling period $T$, which is made explicit by the argument. However, there may be situations when the sampling period $T$ is also subject to design. In this case, we need a metric that is independent of $T$ as well. In this situation, we may use the performance metric to be the limiting value 
\begin{align}
\text{Var}_0 \triangleq \min_T \max_{t\in[0,T]}\text{Var}\left(\theta_t|\bo y\right),  \label{def:var0}
\end{align}
which assumes that we select the sampling period $T$ in a manner that minimizes the worst-case conditional variance for all time. We would consider both the metrics \eqref{def:avar} and \eqref{def:var0} in this paper. 

\section{Main Results}
\subsection{Single snapshot estimation} \label{sec:snapshot}
As motivated earlier, we consider an $N$-sensor network, the collaboration topology of which consists entirely of $Q$-cliques, where $Q$ is a finite integer (see Figure \ref{fig:Qconn}). Let $N=K Q$, which ensures that there is an integral number of such cliques. Let the total energy available in the network be $\mathcal E$, which is finite. We consider the asymptotic limit when the network is large ($N\rightarrow \infty$) but the transmission capacity of the equivalent Multiple-Input-Single-Output (MISO) channel is kept finite. We assume that the random variables $\{\widetilde g_n\}_{n=1}^N$ (which can be thought of as \emph{unnormalized} channel gains) are iid realizations from the pdf $f_{\widetilde g}(\cdot)$ and the channel gains are $ g_n=\frac{1}{\sqrt{N}} \widetilde g_n$ so as to ensure that the transmission capacity remains the same even as the number of nodes increase\footnote{Note that the channel capacity of the equivalent MISO channel is $\frac{1}{2}\log\left(1+\frac{\mathcal E \| \bo g \|^2}{\xi^2} \right)$ and that $\lim_{N\rightarrow \infty} \| \bo g \|^2=\mathbb E\left[ \widetilde g^2 \right]$ from the law of large numbers.}, thereby enabling a fair comparison of networks of various sizes. Without such a scaling, the transmission capacity would increase to infinity (and the resulting distortion would be driven down to zero) as the number of nodes increase, which is a trivial regime to consider. 

Let $J_\textsf{opt}$ and $J_{\textsf{eq}}$ denote the asymptotic limits of the Fisher Information $J_{\bo W}$ corresponding to the optimal \eqref{EA:opt} and equal energy-allocation strategies \eqref{EA:eq} respectively. The following results provide closed form expressions for these limits.

\begin{result:J}[Fisher Information for $Q$-clique topology] \label{result:J:lbl} 
\begin{subequations}
\begin{align}
\mbox{(Optimal EA) }\quad J_\textsf{opt}&=\frac{ \mathcal E }{\eta^2} \frac{\mathbb E\left[ \widetilde g^2 \right]}{\xi^2} \left(1-H_Q \right), \mbox{ and}\label{J:opt} \\
\mbox{(Equal EA) } \quad J_{\textsf{eq}}&=\frac{ \mathcal E }{\eta^2} \frac{ \left( \mathbb E\left[ \widetilde g \right]\right)^2}{\xi^2 } \frac{1}{1+R_Q} \label{J:eq},
\end{align}
\end{subequations}
where $H_Q$ and $R_Q$ are defined as
\begin{subequations}
\begin{align}
H_Q&=\mathbb E\left[\frac{1}{1+\frac{\eta^2}{\sigma^2} \left(h_1^2+\cdots+h_Q^2 \right)} \right], \mbox{ and} \label{HQ}\\
R_Q&=\frac{1}{Q \left( \mathbb E\left[ h \right] \right)^2} \left(\text{Var}\left[ h \right]+\frac{\sigma^2}{\eta^2} \right) \label{RQ},
\end{align}
\end{subequations}
respectively.
\end{result:J}

The proof of Theorem \ref{result:J:lbl} is skipped here due to lack of space and can be found in an extended version of this paper \cite{Kar12CollaborativeArxiv}. A few remarks due to Theorem \ref{result:J:lbl} are in order. 

\emph{Special Cases: } In general, the equal EA scheme is suboptimal, i.e., $J_\textsf{opt} \ge J_{\textsf{eq}}$. However, the two energy allocation schemes are asymptotically equivalent when $\text{Var}[h]$ and $\text{Var}[\widetilde g]$ are both zero, which is the case when the network is homogeneous, i.e., $\bo h= h_0 \bo 1$ and $\tbo g= \widetilde g_0 \bo 1$ (say). For such a network,
\begin{align}
J_\textsf{opt}=J_{\textsf{eq}}=\frac{\mathcal E \widetilde g_0^2}{\xi^2 \eta^2} \frac{1}{1+\frac{1}{Q} \frac{\sigma^2}{\eta^2 h_0^2} }.
\end{align}

\emph{Explicit expressions for Rayleigh distributed gains: } The evaluation of \eqref{J:opt} is, in general, hindered by the computation of $H_Q$, which involves the computation of a $Q$-dimensional integral. However, if the observation gains are Rayleigh\footnote{A Rayleigh distributed random variable $x$ with parameter $\alpha$ has a probability density function $\textsf{Rayleigh}(x;\alpha)=\frac{x}{\alpha^2} \exp\left( -\frac{x^2}{2 \alpha^2} \right)$ for $x\in[0,\infty)$. The first two moments are $\mathbb E\left[ x \right]=\alpha\sqrt{\frac{\pi}{2}}$ and $\mathbb E\left[x^2\right]=2\alpha^2$ respectively.} distributed $f_h(h)=\textsf{Rayleigh}(h;\alpha_h)$, we can show (the derivation is relegated to \cite{Kar12CollaborativeArxiv}) that 
\begin{align} 
H_Q=\frac{(-1)^{Q-1}\lambda^Q \exp (\lambda) \mathcal Ei(\lambda) -\sum_{i=0}^{Q-2} i! (-\lambda)^{Q-1-i}}{(Q-1)!}, \label{HQ:ray}
\end{align}
where $\lambda \triangleq \frac{\sigma^2}{2 \alpha_h^2 \eta^2}$ and $\mathcal Ei(z)\triangleq \int_{z}^\infty \exp(-t)/t \ud t$ is the exponential integral function.  It immediately follows that $H_1=\lambda^Q \exp (\lambda) \mathcal Ei(\lambda)$, which corresponds to the distributed case ($Q=1$), and $H_Q \approx \frac{\lambda}{Q-1}$ for large values of $Q$. In our numerical simulations, we would consider Rayleigh distributed channel and observation gains, for which \eqref{HQ:ray} will be useful.

\subsection*{Simulation results}
Theorem \ref{result:J:lbl} is important since it provides a framework to evaluate the estimation performance for partially connected collaboration topologies. Though \eqref{J:opt} and \eqref{J:eq} are accurate indicators of performance for a structured network consisting only of $Q$-cliques, it is of interest to see how this insight applies for more complicated topologies. Towards that goal, we simulate the  nearest-neighbor and random geometric graph topologies as described in Section \ref{sec:partial}. We consider a network with $N=10^4$ nodes, which is large enough to demonstrate convergent behavior. We consider $\eta^2=1$, $\xi^2=1$ $f_h(h)=\textsf{Rayleigh}(h;1)$, $f_{\widetilde g}(\widetilde g)=\textsf{Rayleigh}(\widetilde g;1)$ and two values of observation noise variance, namely $\sigma^2=1$ and $\sigma^2=2$. The operating energy is fixed at $\mathcal E=0.7$. 
This choice of $\mathcal E$ is made to reflect an operating region where substantial performance gain is possible through spatial collaboration.

\begin{figure}[htb]
\centering
\subfigure[Fixed number of nearest neighbors ]{
\includegraphics[width=\figsza \columnwidth]{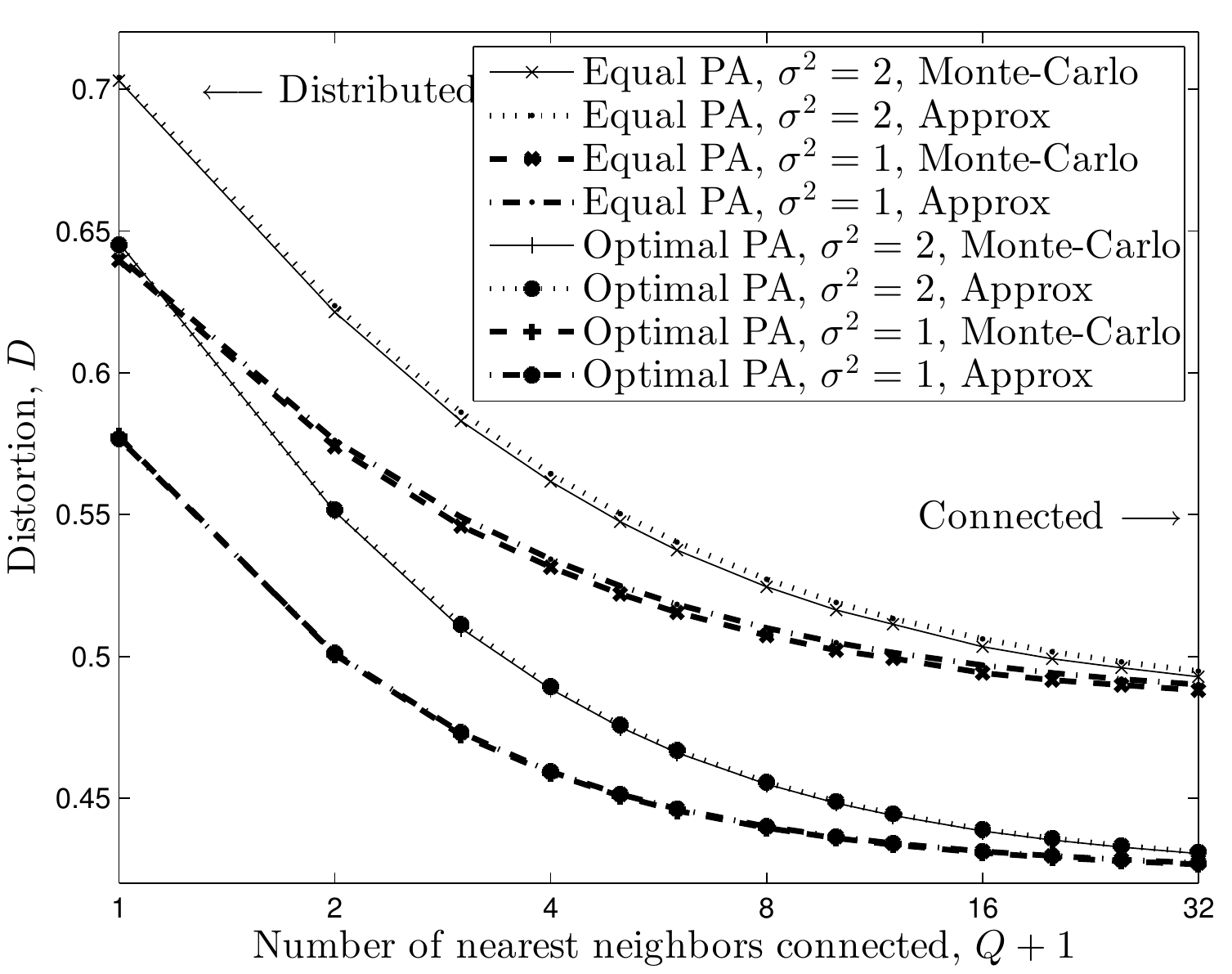}
\label{fig:asymp:nn}
}
\subfigure[Random geometric graph]{
\includegraphics[width=\figsza \columnwidth]{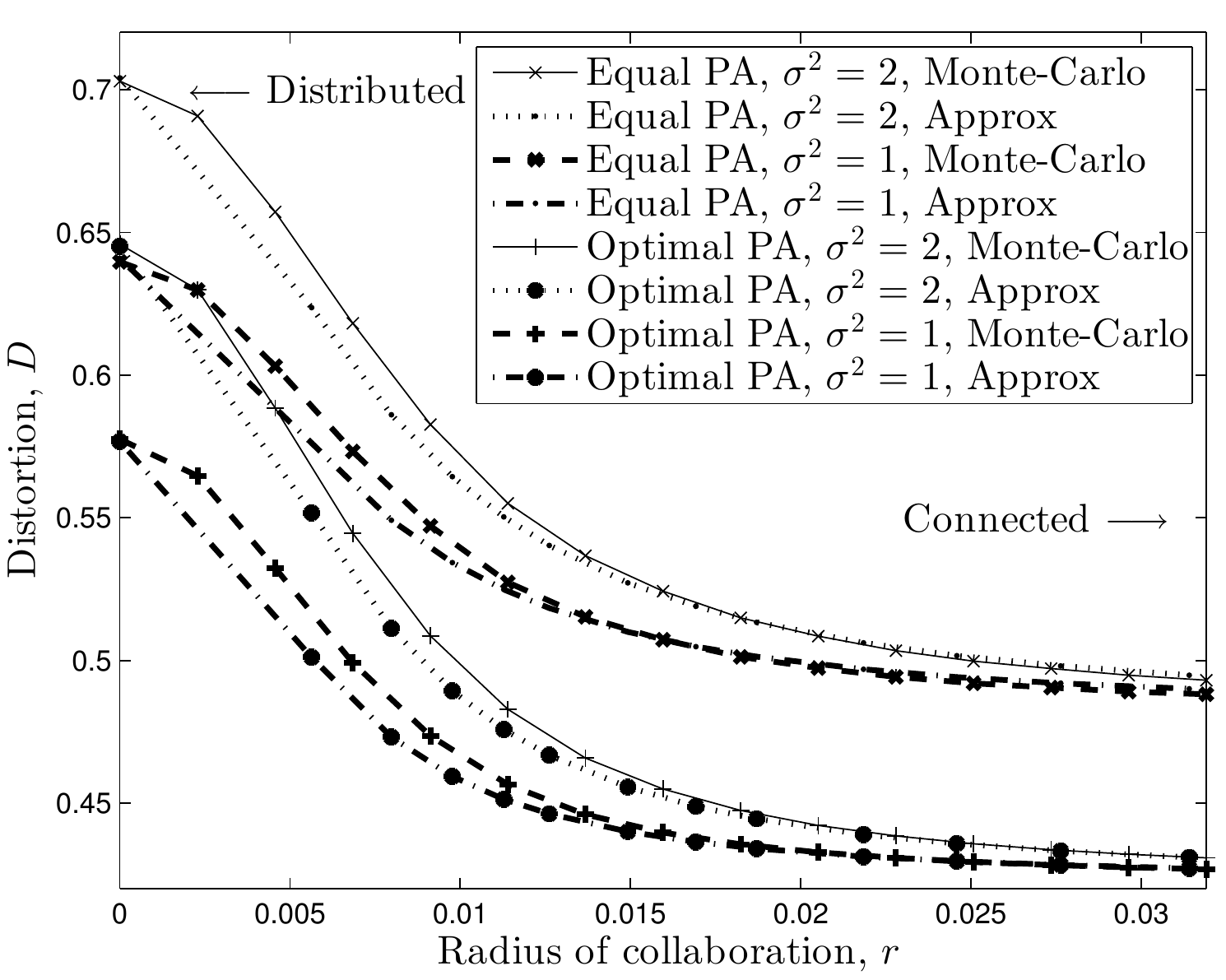}
\label{fig:asymp:rgg}
}
\caption{Energy-constrained estimation with single snapshot spatial sampling.}
\label{fig:asymp}
\end{figure}

Numerical results obtained through Monte-Carlo simulations of both the optimal and equal EA strategies with varying degrees of spatial collaboration are illustrated in Figures \ref{fig:asymp:nn} and \ref{fig:asymp:rgg} for the NN and RGG topologies respectively. For the $(Q-1)$-nearest-neighbor case, the theoretical results corresponding to an equivalent problem with $Q$-cliques are juxtaposed. It is observed from Figure \ref{fig:asymp:nn} that the performance of the two schemes are almost identical. For a random geometric graph, we consider that all the $N$ sensors are randomly spread in a unit square. If the radius of collaboration is $r$, it follows that the expected number of neighbors is approximately $\widetilde Q = N \pi r^2$, and that this approximation is more accurate for large values of $r$. Hence in Figure \ref{fig:asymp:rgg}, the theoretical results corresponding to an equivalent problem with $\widetilde Q$-cliques are juxtaposed. It is observed that the theoretical approximations compare favorably with the Monte-Carlo simulations, although they are less accurate compared to the $(Q-1)$-nearest-neighbor case. 

From the two examples given above, it is observed that only a small number of collaboration links are needed to achieve near-connected performance. In particular, the distortion performance is seen to saturate as early as $Q\gtrapprox 20$, though a fully connected network would imply $Q=N=10^4$ connections per node. This demonstrates the efficacy of spatial collaboration as an approach to enhance estimation performance beyond distributed networks. 

\subsection{Time varying process estimation} \label{sec:OU}
In this subsection, we compute the conditional variance of the OU process given the vector of periodically sampled observations. Towards computing \eqref{var:theta:t}, we begin by describing the matrix $\bo R_{\bo y \bo y'}$ and vector $\bo R_{\theta_t \bo y'}$. The covariance matrix of the sampled parameter values $\bo \theta$ takes the shape of the well known stationary matrix  (e.g., \cite{Niu10}, \cite{Kar12}),
\begin{align}
\mathbb{E}[\bo \theta \bo \theta'] &= \eta^2\bo C, \; \bo C \triangleq \begin{bmatrix}
1         & \rho  & \rho^2 & \hdots   &    \cdot \\
\rho     & 1      &  \ddots & \ddots    &   \vdots \\
\rho^2 & \ddots  & \ddots & \ddots  &   \rho^2 \\
\vdots  & \ddots  & \ddots & 1         &  \rho \\
\cdot   & \hdots   & \rho^2  & \rho     & 1    \\
\end{bmatrix}, \label{R:theta}
\end{align}
where $\rho \triangleq e^{-T/\tau}$. The structured matrix $\bo C$ in Equation \eqref{R:theta}, is often referred to as the Kac--Murdock--Szeg$\ddot{o}$ matrix in the literature. From the additive model \eqref{y:theta:v}, it follows that
\begin{align}
\bo R_{\bo y \bo y'}=\left( \bo I+ \eta^2 J \bo C \right)/J.
\end{align} 
Similarly, the following expression for $\bo R_{\theta_t \bo y'}$ follows directly from the definition $\varrho = e^{-t/\tau}$, where $t\in[0,T]$,
\begin{align}
\mathbb{E}[\theta_t \bo y'] &= \eta^2 \left[ \ldots, \rho^2\varrho,\rho\varrho,\varrho, \rho/\varrho,\rho^2/\varrho,\ldots\right].
\end{align}
With the help of the above descriptions of $\bo R_{\bo y \bo y'}$ and $\bo R_{\theta_t \bo y'}$, computing \eqref{var:theta:t} involves inverting the matrix $\bo I+ \eta^2 J \bo C$ (the asymptotic closed form expression for such an inverse was introduced in \cite{Kar12}) followed by a quadratic product. The resulting algebra is involved but straightforward. We relegate details of the derivation to \cite{Kar12CollaborativeArxiv} and state the result below.

\begin{result:var:theta}[Variance of OU process estimates] \label{result:var:theta:lbl} 
\begin{align}\begin{split}
\text{Var}\left(\theta_t|\bo y\right)&=\frac{\eta^2\left[1+\eta^2 J\rho' \left\{ 1- \left(\frac{\varrho-\rho/\varrho}{1-\rho} \right)^2 \right\} \right]}{ \sqrt{ \left( \eta^2 J+\rho' \right) \left( \eta^2 J+1/\rho' \right) } }, \\
\mbox{where } \rho' &\triangleq \frac{1-\rho}{1+\rho},  \rho=e^{-T/\tau}, \varrho=e^{-t/\tau},  t\in[0,T].
\label{res:var:theta}
\end{split}\end{align} 
\end{result:var:theta}
Equation \eqref{res:var:theta} provides the closed form estimation variance of an OU process at any instant $t\in[0,T]$, provided that power constrained noisy samples are observed periodically with period $T$. In addition to $\rho$, the quantity $J$ also depends on the sampling period $T$ through the energy-FI (Fisher Information) relation $J=\frac{c PT}{\eta^2}$, which follows from Theorem \ref{result:J:lbl} by using $\mathcal E=P T$ and defining 
\begin{equation}
c \triangleq \begin{cases}
\frac{\mathbb E\left[ \widetilde g^2 \right] \left(1-H_Q \right)}{\xi^2} & \mbox{ for Optimal EA},\\
\frac{\mathbb (E\left[ \widetilde g \right])^2}{\xi^2 \left(1+R_Q \right)} & \mbox{ for Equal EA}.
\end{cases} \label{def:c}
\end{equation}
In the following discussions, we illustrate the power-constrained estimation of an OU process and show how the instantaneous variance \eqref{res:var:theta} can be used to compute other performance measures described in Section \ref{sec:prob:form:OU}, namely a) average variance, $\text{Avar}(T)$ and b) min-max performance limit, $\text{Var}_0$.

\subsection*{Simulation results} 
\begin{figure}[htb]
\centering
\subfigure[Effect of sampling period on aggregate noise and subsequent filtering]{
\includegraphics[width=\figsza \columnwidth]{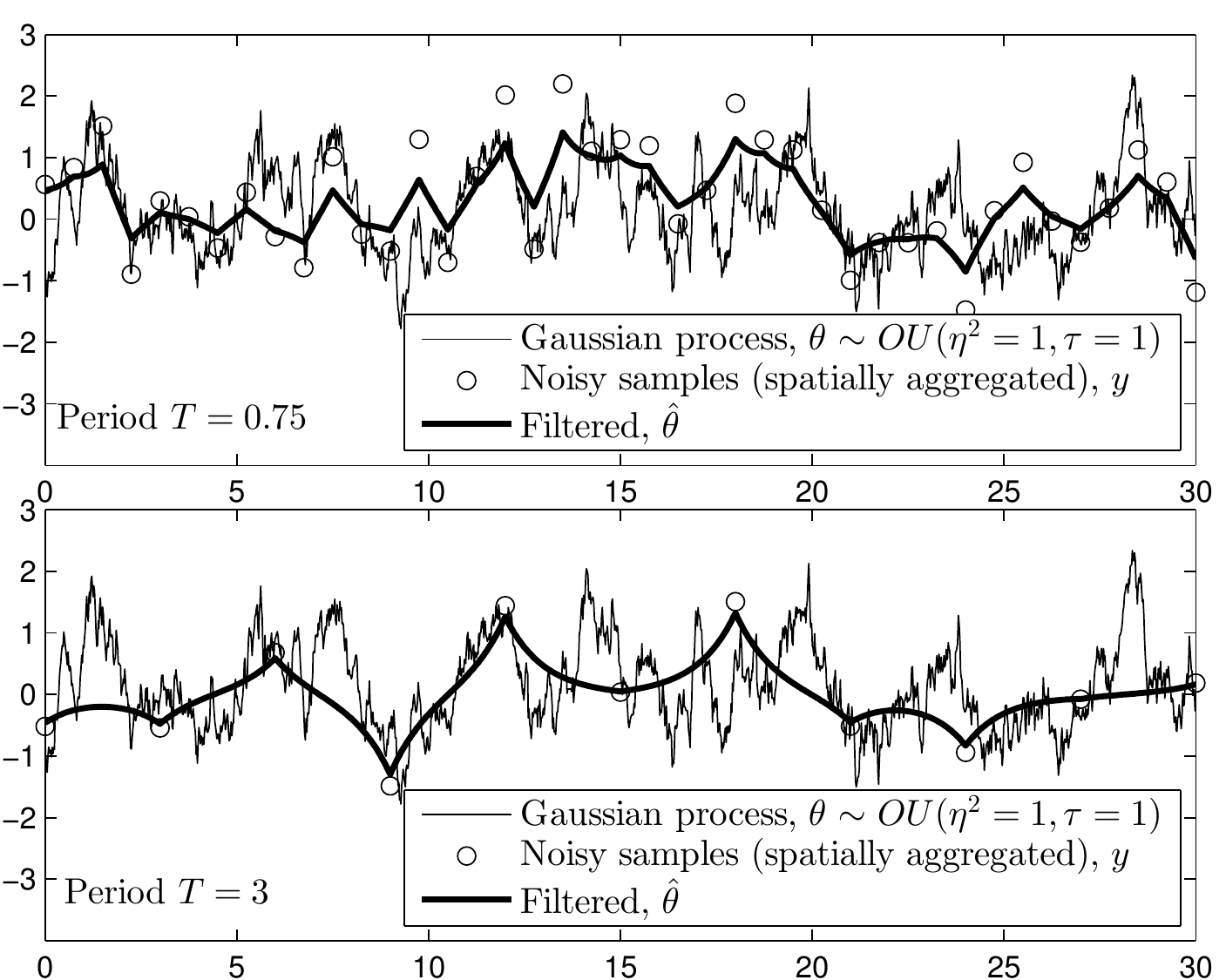}
\label{fig:OUproc:eg}
}
\subfigure[Instantaneous variance for various sampling periods]{
\includegraphics[width=\figsza \columnwidth]{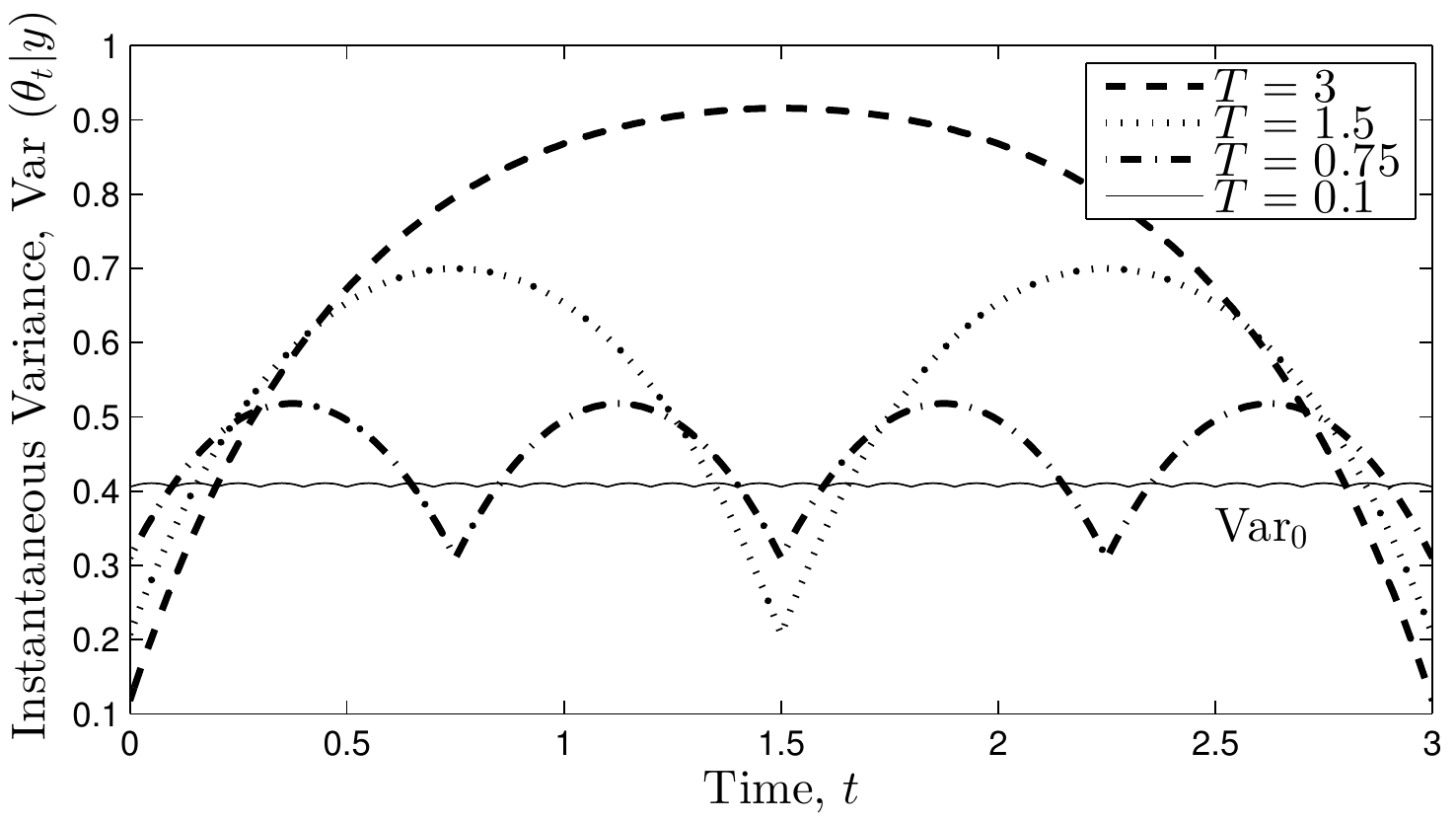}
\label{fig:OUproc:var}
}
\caption{Power-constrained estimation of OU process }
\vspace{-0.2in}
\label{fig:OUproc}
\end{figure}

In Figure \ref{fig:OUproc:eg}, we visualize the estimation of an OU process with stationary variance $\eta^2=1$ and covariance drop-off parameter $\tau=1$s. We simulate a total duration of $T_\textsf{obs}=30$s, during which we consider sampling the same process using two different sampling periods, $T=0.75$s (top) and $T=3$s (bottom).  The sampling noise sequence $\{ v_{kT} \}$, which is an abstraction of the spatial data aggregation, is simulated as independent zero-mean Gaussian random variables with variance $\frac{1}{2.5 T}$ for the two different sampling periods. The inverse relation $\text{Var}\left(v_{kT}\right) \propto \frac{1}{T}$ is due to the fact that $v_{kT}$ represents a noise with variance $\frac{1}{J}$ and the Fisher Information $J=\frac{c P T}{\eta^2}$ as per Theorem \ref{result:J:lbl}. The constant $2.5$ (which represents the quantity $\frac{cP}{\eta^2}$) was chosen so as to produce a visible contrast between the sampling errors corresponding to the chosen sampling periods. As can be seen in Figure \ref{fig:OUproc:eg}, the samples are obtained almost without any noise for $T=3$s (bottom). The circles representing noisy samples align almost on top of the thin line that represents the path of the OU process. The samples are, however, significantly noisy for $T=0.75$s (top), since less energy is available per sampling duration. This is evidenced by the circles lying significantly distant from the OU process path. The filtered estimates $\widehat \theta_t=\mathbb E\left[ \theta_t|\{ y_{kT} \} \right]$ are obtained by applying \eqref{E:theta:t} and are shown by the bold lines. When compared visually, the filtered estimates appear more accurate in the case of smaller sampling period (top). This observation is justified by plotting the steady state variance, as obtained from Theorem \ref{result:var:theta:lbl}, in Figure \ref{fig:OUproc:var} for various sampling periods $T=\{0.1,0.75,1.5,3\}$. Though the best-case variance (occurring at $t=kT$) is higher for smaller sampling periods, the worst-case variance (occurring at $t=(k+0.5)T$) goes down as the OU process is sampled more frequently. Because the power is kept constant, the variance converges to a finite value (rather than vanishing) for small values of $T$. Since the gap between the worst-case and best-case scenarios reduces with $T$, the limiting variance ($\approx 0.4$, annotated as $\text{Var}_0$) is flat with respect to time. From Equation \eqref{res:var:theta}, the limiting variance can be derived precisely to be 
\begin{align}
\text{Var}_0\triangleq \lim_{T\rightarrow 0} \text{Var}\left(\theta_t|\bo y\right) = \frac{\eta^2}{\sqrt{1+2 P\tau c}}, \label{var0}
\end{align}
which also means that $\text{Var}_0$ trivially satisfies
\begin{align}
\min_T \max_{t\in[0,T]}\text{Var}\left(\theta_t|\bo y\right)=\text{Var}_0,
\end{align}
thereby answering the question of min-max performance limit as posed earlier in Equation \eqref{def:var0}. The following result is obtained by substituting in \eqref{var0} the value of constant $c$ (see \eqref{def:c}), thereby stating explicitly how the performance limit depends on channel conditions and collaboration topology.

\begin{result:var0}[Min-max performance limits] \label{result:var0:lbl} 
\begin{equation}
\text{Var}_0 = \begin{cases}
\eta^2 \bigg/ \sqrt{1+ \frac{2P \tau \mathbb E\left[ \widetilde g^2 \right] \left(1-H_Q \right)}{\xi^2}} & \mbox{ for Optimal EA},\\
\eta^2 \bigg/ \sqrt{1+ \frac{2P \tau (\mathbb E\left[ \widetilde g \right])^2}{\xi^2 \left(1+R_Q \right)}} & \mbox{ for Equal EA}.
\end{cases} \label{def:c}
\end{equation}
\end{result:var0}

In addition to the instantaneous variance and min-max performance limits, one may also be interested in the average variance $\text{Avar}(T)$ as defined by Equation \eqref{def:avar} in Section \ref{sec:prob:form:OU}. The average variance is obtained by integrating \eqref{res:var:theta} over $t\in[0,T]$. Since $\varrho=e^{-t/\tau}$ is the only variable in \eqref{res:var:theta} that depends on $t$, we obtain
\begin{align}\begin{split}
\text{Avar}(T)&=\frac{\eta^2 \left[1+\eta^2 J \rho' \left\{ 1-\frac{ \mathcal I_T  }{ \left( 1-\rho \right)^2}  \right\} \right]}{ \sqrt{ \left( \eta^2 J+\rho' \right) \left( \eta^2 J+1/\rho' \right) } }, \\
\mbox{where } \mathcal I_T &\triangleq \frac{1}{T}\int_0^T \left( \varrho -\rho/\varrho \right)^2 \ud t=\frac{1-\rho^2}{T/\tau}-2\rho.
\end{split}\end{align}

\begin{figure}[htb]
\centering
\subfigure[Fixed number of nearest neighbors]{
\includegraphics[width=\figsza \columnwidth]{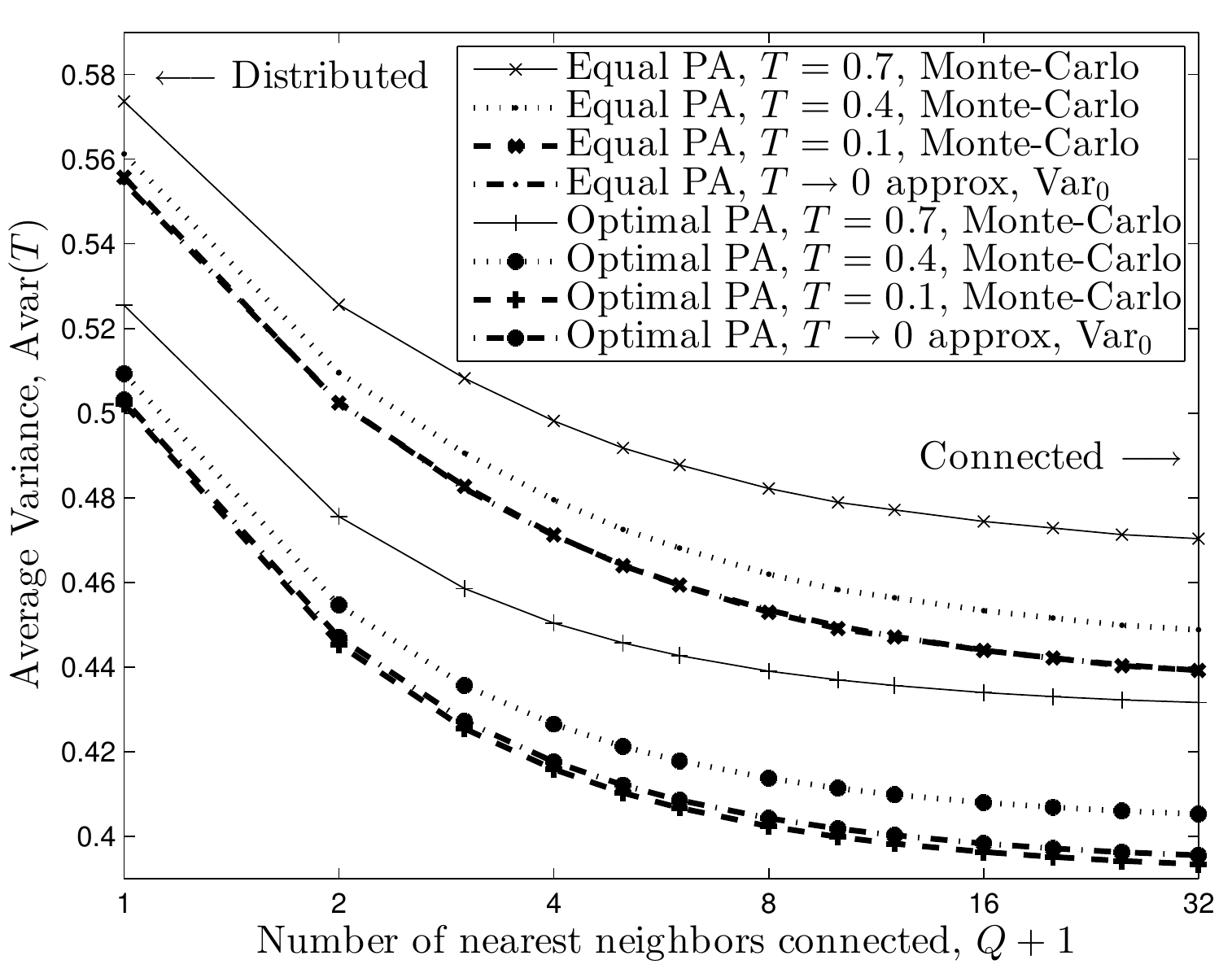}
\label{fig:asymp:nn:pow}
}
\subfigure[Random geometric graph]{
\includegraphics[width=\figsza \columnwidth]{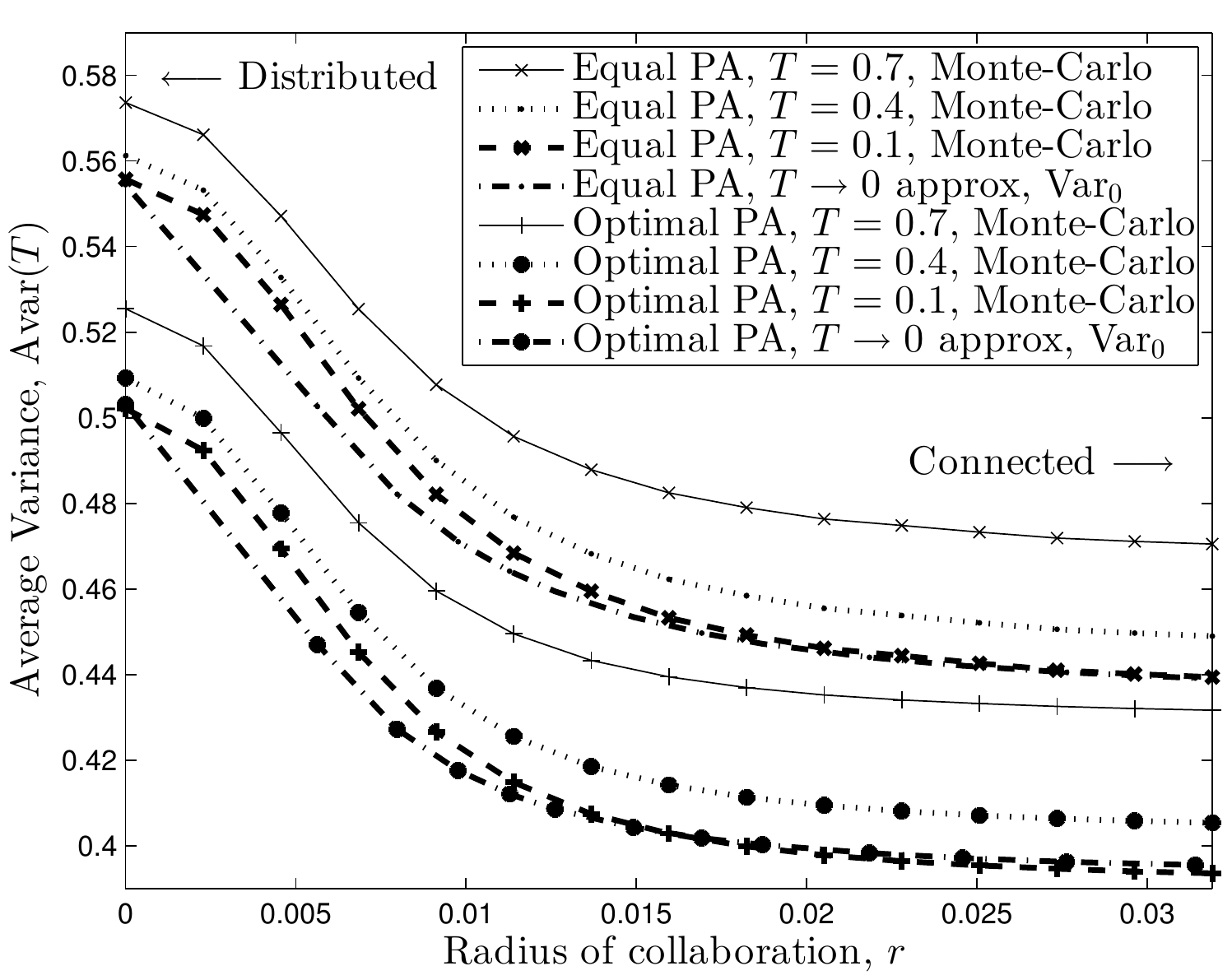}
\label{fig:asymp:rgg:pow}
}
\caption{Power-constrained estimation of OU process - Average variance}
\vspace{-0.3in}
\label{fig:asymp:pow}
\end{figure}

We use average variance as the performance metric in the following simulation, in which we consider both the aspects discussed in this paper, namely the spatial aggregation procedure and the temporal dynamics. The simulation settings are similar to those considered in Section \ref{sec:snapshot}, which we repeat here for the sake of completeness. The sensor network comprises of $N=10^4$ nodes. We consider $\eta^2=1$, $\xi^2=1$, $\sigma^2=1$, $f_h(h)=\textsf{Rayleigh}(h;1)$ and $f_g(g)=\textsf{Rayleigh}(g;1)$. The exponential drop-off parameter is set to $\tau=1$s and an observation duration of $T_\textsf{obs}=30$s is considered, which is large enough to demonstrate steady state behavior. The observation duration is discretized into $M=1600$ instants for generating the OU process sequence using a first-order autoregressive model. The average variance is obtained as the mean of the deviations from all $M$ estimates. Three values of sampling period are considered for simulation, namely $T=\{0.7,0.4,0.1\}$. The limiting value when $T\rightarrow 0$, $\text{Var}_0$, is also shown on all the graphs. The operating power is chosen as $P=1.4$ to reflect an operating region where substantial performance gain is possible through spatial collaboration. Both NN (Figure \ref{fig:asymp:nn:pow}) and RGG (Figure \ref{fig:asymp:rgg:pow}) topologies are considered to show the applicability of the $Q$-clique results to practical collaboration scenarios. As earlier, both the equal-EA and optimal-EA spatial energy allocation strategies are simulated. The results in Figure \ref{fig:asymp:pow} show that the average variance decreases with $T$. Though we have not rigorously proved that $\text{Avar}(T)$ is monotonically decreasing in $T$, this assertion can be visually verified from Figure \ref{fig:OUproc:var}, by comparing the area under the curves for any two sampling periods ($T=3$ and $T=1.5$, say).  This observation coupled with the min-max property of $\text{Var}_0$ leads to the conclusion that an OU process should be sampled as frequently as possible, even if that implies that less energy is available per sampling period (resulting in more noisy samples). However, this assertion is based on the assumption that the sampling noise is temporally white. In practical situations, the sampling errors may become correlated if the samples are obtained too frequently, and caution must be exercised to make sure that the temporal independence assumption is valid.

\section{Conclusion}
In this paper, we have considered the linear coherent estimation problem in wireless sensor networks and investigated two key aspects. First, we have provided an asymptotic analysis of the single-snapshot estimation problem when the collaboration topology is only partially connected.  We achieve this by obtaining the solutions for a family of structured networks and then using those solutions to approximately predict the performance of more sophisticated networks using geometric arguments. Second, we have extended the problem formulation towards the estimation of a time varying signal. In particular, we have derived the instantaneous, average and worst case performance metrics when the signal is modeled as a Gaussian random process with exponential covariance. Both these aspects were investigated under the assumption of spatial and temporal independence among the measurement and channel noise samples. In the future, we plan to relax this assumption and observe the effect of spatial and temporal correlation on the estimation performance.

\bibliographystyle{IEEEtran}
\bibliography{thesis}

\begin{thebibliography}{10}
\providecommand{\url}[1]{#1}
\csname url@samestyle\endcsname
\providecommand{\newblock}{\relax}
\providecommand{\bibinfo}[2]{#2}
\providecommand{\BIBentrySTDinterwordspacing}{\spaceskip=0pt\relax}
\providecommand{\BIBentryALTinterwordstretchfactor}{4}
\providecommand{\BIBentryALTinterwordspacing}{\spaceskip=\fontdimen2\font plus
\BIBentryALTinterwordstretchfactor\fontdimen3\font minus
  \fontdimen4\font\relax}
\providecommand{\BIBforeignlanguage}[2]{{%
\expandafter\ifx\csname l@#1\endcsname\relax
\typeout{** WARNING: IEEEtran.bst: No hyphenation pattern has been}%
\typeout{** loaded for the language `#1'. Using the pattern for}%
\typeout{** the default language instead.}%
\else
\language=\csname l@#1\endcsname
\fi
#2}}
\providecommand{\BIBdecl}{\relax}
\BIBdecl

\bibitem{Rib06}
A.~Ribeiro and G.~B. Giannakis, ``Bandwidth-constrained distributed estimation
  for wireless sensor networks-{Part I: Gaussian} case,'' \emph{Signal
  Processing, IEEE Transactions on}, vol.~54, no.~3, pp. 1131--1143, 2006.

\bibitem{Cui07}
S.~Cui, J.-J. Xiao, A.~Goldsmith, Z.-Q. Luo, and H.~Poor, ``Estimation
  diversity and energy efficiency in distributed sensing,'' \emph{Signal
  Processing, IEEE Transactions on}, vol.~55, no.~9, pp. 4683--4695, Sept.
  2007.

\bibitem{Xiao08}
J.-J. Xiao, S.~Cui, Z.-Q. Luo, and A.~Goldsmith, ``Linear coherent
  decentralized estimation,'' \emph{Signal Processing, IEEE Transactions on},
  vol.~56, no.~2, pp. 757--770, Feb. 2008.

\bibitem{KarTSP12}
S.~Kar, H.~Chen, and P.~K. Varshney, ``Optimal identical binary quantizer
  design for distributed estimation,'' \emph{Signal Processing, IEEE
  Transactions on}, vol.~60, no.~7, pp. 3896--3901, July 2012.

\bibitem{Fang09}
J.~Fang and H.~Li, ``Power constrained distributed estimation with
  cluster-based sensor collaboration,'' \emph{Wireless Communications, IEEE
  Transactions on}, vol.~8, no.~7, pp. 3822--3832, July 2009.

\bibitem{KarISIT12}
S.~Kar and P.~K. Varshney, ``On linear coherent estimation with spatial
  collaboration,'' in \emph{Information Theory Proceedings (ISIT), 2012 IEEE
  International Symposium on}, July 2012, pp. 1448--1452.

\bibitem{Gastpar03}
M.~Gastpar, B.~Rimoldi, and M.~Vetterli, ``To code, or not to code: {Lossy}
  source-channel communication revisited,'' \emph{Information Theory, IEEE
  Transactions on}, vol.~49, no.~5, pp. 1147--1158, May 2003.

\bibitem{Kay93}
S.~M. Kay, \emph{Fundamentals of Statistical Signal Processing: Estimation
  Theory}.\hskip 1em plus 0.5em minus 0.4em\relax Englewood Cliffs, NJ:
  Prentice Hall, 1993.

\bibitem{Freris10}
N.~Freris, H.~Kowshik, and P.~Kumar, ``Fundamentals of large sensor networks:
  Connectivity, capacity, clocks, and computation,'' \emph{Proceedings of the
  IEEE}, vol.~98, no.~11, pp. 1828--1846, Nov. 2010.

\bibitem{Niu10}
R.~Niu and P.~K. Varshney, ``Sampling schemes for sequential detection with
  dependent observations,'' \emph{Signal Processing, IEEE Transactions on},
  vol.~58, no.~3, pp. 1469--1481, Mar. 2010.

\bibitem{Kar12CollaborativeArxiv}
\BIBentryALTinterwordspacing
S.~Kar and P.~K. Varshney, ``Controlled collaboration for linear coherent
  estimation in wireless sensor networks,'' \emph{{arXiv:1210.1624}}. [Online].
  Available: \url{http://arxiv.org/abs/1210.1624}
\BIBentrySTDinterwordspacing

\bibitem{Kar12}
S.~Kar, P.~K. Varshney, and M.~Palaniswami, ``{Cram\'er-Rao} bounds for
  polynomial signal estimation using sensors with {AR(1)} drift,'' \emph{Signal
  Processing, IEEE Transactions on}, vol.~60, no.~10, pp. 5494--5507, Oct.
  2012.

\end{thebibliography}

\end{document}